\documentclass{ws-ijqi}
  \usepackage{amsfonts,amssymb,amsmath}
  \usepackage{latexsym}

\newcommand{\eq}{\begin{equation}}
\newcommand{\en}{\end{equation}}
\newcommand{\eqa}{\begin{eqnarray}}
\newcommand{\ena}{\end{eqnarray}}

\begin{document}

\markboth{Yong Zhang, Louis H. Kauffman, Mo-Lin Ge} {Universal
Quantum Gate, Yang--Baxterization and Hamiltonian}

\catchline{}{}{}{}{}

\title{Universal Quantum Gate, Yang--Baxterization and Hamiltonian}

\author{Yong Zhang}

\address{Institute of Theoretical Physics, Chinese Academy of Sciences \\
P. O. Box 2735, Beijing 100080, P. R. China\\
yong@itp.ac.cn}

\author{Louis H. Kauffman}

\address{Department of Mathematics, Statistics and Computer
 Science\\ University of Illinois at Chicago\\  851 South Morgan Street
Chicago, IL, 60607-7045\\
kauffman@uic.edu}

\author{Mo-Lin Ge}

\address{Nankai Institute of Mathematics, Nankai University \\
Tianjin 300071, P. R. China \\
geml@nankai.edu.cn}

\maketitle

\begin{history}
\received{29 December 2004}
\end{history}

\begin{abstract}
It is fundamental to view unitary braiding operators describing
topological entanglements as universal quantum gates for quantum
computation. This paper derives a unitary solution of the quantum
Yang--Baxter equation via Yang--Baxterization and constructs the
Hamiltonian responsible for the time-evolution of the unitary
braiding operator.
\end{abstract}



\section{Introduction}

There are natural relationships between quantum entanglements
\cite{nielsen} and topological entanglements \cite{kauffman0}.
Topology studies global relationships in spaces, and how one space
can be placed within another, such as knotting and linking of
curves in three-dimensional space. One way to study topological
entanglement and quantum entanglement is to try making direct
correspondences between patterns of topological linking and
entangled quantum states. The approach of this kind was initiated
by Aravind \cite{aravind}, suggesting that observation of a link
would be modelled by deleting one component of the link. But this
correspondence property of quantum states and topological links is
not basis independent \cite{aravind}.

A deeper method (we believe) is to consider unitary gates $\check R$
that are both universal for quantum computation and are also
solutions to the condition for topological braiding. Such $\check
R$-matrices are unitary solutions to the quantum Yang--Baxter
equation (QYBE), and can be used as universal quantum gates in the
sense discussed in this paper. We are then in a position to compare
the topological and quantum properties of these transformations. In
this way, we can study the apparently complex relationship among
topological entanglement, quantum entanglement, and quantum
computational universality. This way has been explored in a series
of papers, see Refs. \refcite{kauffman1}--\refcite{dye} for more
details.

The present paper is an outgrowth of Ref. \refcite{kauffman8}. It
derives the unitary solution of the QYBE via Yang--Baxterization and
explores the corresponding dynamical evolution of quantum states.
The solutions to the QYBE that we derive by Yang--Baxterization
contain a spectral parameter, and hence do not, except in special
cases, give representations of the Artin braid group. These new
solutions are unitary, and they do give useful quantum gates. Thus
we show in this paper that the full physical subject of solutions to
the Yang--Baxter equation (including the spectral parameter) is of
interest for quantum computing and quantum information theory.

The plan of the paper is organized as follows. In the second
section, we relate a unitary solution of the braid relation to a
universal quantum gate. In the third section, we obtain a unitary
solution of the QYBE via Yang--Baxterization acting on the solution
of the braid relation. In the fourth section, we construct the
Hamiltonian responsible for the time-evolution of the braiding
operator and write down the Schr{\"o}dinger equation. In the fifth
section, we generate the Bell states by the $\check{R}$-matrix
acting on unentangled states and construct the CNOT gate via single
qubit transformations acting on the $\check{R}$-matrix. At last,
remarks on further research are made.

 \section{The QYBE and universal quantum gate}

 In this article, the braid group representation (BGR) $b$-matrix
 \cite{kauffman2} and QYBE solution
 $\check{R}$-matrix, see Refs. \refcite{yang}--\refcite{jimbo}, are $ n^2
 \times n^2$ matrices acting on $ V\otimes V$ where $V$ is an
 $n$-dimensional vector space. As $b$ and $\check{R}$ act on the
 tensor product $V_i\otimes V_{i+1}$, we denote them by $b_i$ and
 $\check{R}_i$, respectively.

 The BGR $b$-matrix has to satisfy the braid  relation \eq \label{bgr}
 b_i\,b_{i+1}\,b_i=b_{i+1}\,b_i\,b_{i+1}, \en while the QYBE has
 the form \eq \label{qybe}
 \check{R}_i(x)\,\check{R}_{i+1}(xy)\,\check{R}_i(y)=
 \check{R}_{i+1}\,\check{R}_i(xy)\,\check{R}_{i+1}(x)\en with the
 asymptotic condition \eq \check{R}(x=0)=b, \en and $x$ called the
 spectral parameter. From these two equations both $b$ and
 $\check{R}(x)$ are fixed up to an overall scalar factor.

A two-qubit gate $G$ is a unitary  linear mapping $G:V \otimes V
\longrightarrow V$ where $V$ is a two complex dimensional vector
space. We say that the gate $G$ is { universal for quantum
computation} (or just universal) if $G$ together with local unitary
transformations (unitary transformations from $V$ to $V$) generates
all unitary transformations of the complex vector space of dimension
$2^{n}$ to itself. It is well-known \cite{nielsen} that the CNOT
gate is a universal gate. \bigbreak

\noindent A gate $G$, as above, is said to be {\it entangling} if
there is a vector
$$| \alpha \beta \rangle = | \alpha \rangle \otimes | \beta \rangle \in V \otimes V$$
such that $G | \alpha \beta \rangle$ is not decomposable as a
tensor product of two qubits. Under these circumstances, one says
that $G | \alpha \beta \rangle$ is  entangled. \bigbreak

\noindent In Refs. \refcite{N}--\refcite{BB}, the Brylinskis give a
general criterion of $G$ to be universal. They prove that a
two-qubit gate $G$ is universal if and only if it is {\em
entangling}. \bigbreak

\noindent In Ref. \refcite{kauffman8},  Kauffman and Lomonaco prove
the following result.

\bigbreak

\noindent {\bf Theorem 1}. {\it Let \eq \label{theorem} \check{R} =
\left(
\begin{array}{cccc}
1/\sqrt{2} & 0 & 0  & 1/\sqrt{2}\\
0  & 1/\sqrt{2} & -1/\sqrt{2} & 0\\
0 & 1/\sqrt{2} & 1/\sqrt{2} & 0 \\
-1/\sqrt{2} & 0 & 0 & 1/\sqrt{2}\\
\end{array} \right)\en be the unitary solution to the braid relation (\ref{bgr}).
 Then $\check{R}$ is a universal gate. The proof below
 (repeated from Ref. \refcite{kauffman8}) gives a specific expression for the CNOT gate
 in terms of $\check{R}.$} \bigbreak

\noindent {\bf Proof.} This result follows at once from the
Brylinksis' Theorem \cite{BB}, since $\check{R}$ is highly
entangling. For a direct computational proof, it suffices to show
that the CNOT gate can be generated from $\check{R}$ and local
unitary transformations. Let \eqa \label{delta} \alpha &=& \left(
\begin{array}{cc}
1/\sqrt{2} & 1/\sqrt{2}\\
1/\sqrt{2} & -1/\sqrt{2}\\
\end{array} \right), \qquad
\beta = \left( \begin{array}{cc}
-1/\sqrt{2} & 1/\sqrt{2}\\
i/\sqrt{2} & i/\sqrt{2}\\
\end{array} \right) \nonumber\\
\gamma &=& \left( \begin{array}{cc}
1/\sqrt{2} & i/\sqrt{2}\\
1/\sqrt{2} & -i/\sqrt{2}\\
\end{array} \right),\qquad
\delta = \left( \begin{array}{cc}
1 & 0\\
0 & i\\
\end{array} \right).\ena  Let $M= \alpha \otimes \beta$ and $N= -\gamma \otimes \delta.$
Then it is straightforward to verify that
$$\textrm{CNOT} = M\cdot \check{R} \cdot N.$$ This completes the proof. $\hfill \Box $
\bigbreak

\section{The unitary $\check{R}(x)$-matrix via Yang--Baxterization}

  The QYBE (\ref{qybe}) solution $\check{R}$-matrices usually depend on the deformation
  parameter $q$ and the spectral parameter $x$. With two such parameters, there
  exist two approaches to solving the QYBE (\ref{qybe}).
  Taking the limit of $x\to 0$ leads to the braid relation (\ref{bgr}) from
  the QYBE (\ref{qybe}) and the BGR $b$-matrix from the $\check{R}$-matrix. Concerning relations
  between the BGR and $x$-dependent solutions of the QYBE (\ref{qybe}), we either reduce a known
  $\check{R}(x)$-matrix to a BGR $b$-matrix, see Refs. \refcite{wadati1}--\refcite{turaev}, or construct
  a $\check{R}(x)$-matrix from a given BGR $b$-matrix. Such a construction is called
  Yang--Baxterization.

  In knot theory, these solutions were first studied by Jones
  \cite{jones} and Turaev \cite{turaev} for the BGR satisfying
  the Heck algebra relations and for the BGR satisfying the
  Birman--Wenzl algebra relations (corresponding to the Kauffman two-variable
  polynomial \cite{kauffman0}). Later the more general cases with a BGR having
  three or four unequal eigenvalues were considered in Refs. \refcite{molin1}--\refcite{molin4},
  including all known trigonometric solutions to the QYBE. Also
  Yang--Baxterization of non-standard BGR $b$-matrix has been
  discussed in Refs. \refcite{sogo}--\refcite{schultz1}.

  In this section, we apply Yang--Baxterization to derive a unitary
  $\check{R}(x)$-matrix so that it is possible to find out the
  Hamiltonian controlling the evolution of quantum entangled
  states. As an example, we will present a solution of the
  BGR for the eight-vertex model and its corresponding unitary
  $\check{R}$-matrix via Yang--Baxterization.

  In terms of non-vanishing Boltzman weights $w_1$, $\cdots$, $w_8$, the BGR $b$-matrix
  of the eight-vertex model assumes the form
  \eq b=\left(\begin{array}{cccc}
  w_1 & 0 & 0 & w_7 \\
  0 & w_5 & w_3 & 0 \\
  0 & w_4 & w_6 & 0 \\
  w_8 & 0 & 0 & w_2
  \end{array}\right).\en
 Choosing suitable Boltzman weights leads to solutions of the
 braid relation (\ref{bgr}).

Setting $w_1=w_2=w_5=w_6$ gives us $w_1^2=w_3^2=w_4^2$ and
$w_3^2+w_7 w_8=0$. In the case of $w_3\neq w_4$,  we have
$w_3=-w_4$ and $w_1=\pm w_3$. The BGR $b$-matrix has the form \eq
b_\pm=\left(\begin{array}{cccc}
w_1 & 0 & 0 & w_7 \\
0 & w_1 & \pm w_1 & 0 \\
0 & \mp w_1 & w_1 & 0 \\
-\frac {w_1^2} {w_7} & 0 & 0 & w_1
\end{array}\right)  \Longleftrightarrow
\left(\begin{array}{cccc}
1 & 0 & 0 & q \\
0 & 1 & \pm 1 & 0 \\
0 & \mp 1 & 1 & 0 \\
-q^{-1} & 0 & 0 & 1
\end{array}\right).\en

It has two eigenvalues $\lambda_1=1-i$ and $\lambda_2=1+i$.  The
corresponding $\check{R}(x)$-matrix via Yang--Baxterization is
obtained to be \eqa  \label{belltype1} \check{R}_\pm(x) &=&
 b_\pm+x\,\lambda_1\lambda_2 b_\pm^{-1} \nonumber\\
&=& \left(\begin{array}{cccc}
1+x & 0 & 0 & q(1-x) \\
0 & 1+x & \pm(1-x) & 0 \\
0 & \mp(1-x) & 1+x & 0 \\
-q^{-1} (1-x) & 0 & 0 & 1+x \end{array}\right).  \ena

Assume the spectral parameter $x$ and the deformation parameter
$q$ to be complex numbers. The unitarity condition \eq
\check{R}_\pm(x)\,\check{R}_\pm^\dag(x)=\check{R}_\pm^\dag(x)\,\check{R}_\pm(x)\propto
\rho_\pm 1\!\! 1 \en leads to the following equations \eq \left\{
\begin{array}{ccc}
\|1+x\|^2+\|q\|^2 \|1-x\|^2 &=& \rho_\pm \\
\|1+x\|^2+\frac 1 {\|q\|^2} \|1-x\|^2 &=& \rho_\pm\\
\|1+x\|^2+\|1-x\|^2 &=& \rho_\pm\\
(1-x)(1+\bar x )-(1+x) (1-\bar x)&=& 0\\
-q^{-1} (1-x)(1+\bar x )+\bar q\,(1+x) (1-\bar x)&=&
0\end{array}\right. \en which specify $x$ real and $q$ living at a
unit circle.

Introducing the new variables of angles $\theta$ and $\varphi$ as
follows \eq \label{belltypetran} \cos\theta=\frac 1
{\sqrt{1+x^2}}, \qquad
 \sin\theta=\frac x {\sqrt{1+x^2}}, \qquad
 q=e^{-i\varphi},
 \en
we represent the $\check{R}_\pm(x)$-matrix in a new form \eq
\label{bellrmatrix} \check{R}_\pm(\theta)=\cos\theta\,\,
b_\pm(\varphi)+\sin\theta\,\, (b_\pm)^{-1}(\varphi)\en in which the
BGR $b_\pm(\varphi)$-matrix is given by \eq \label{bgrphi}
 b_\pm(\varphi)=\frac 1 {\sqrt
2}\left(\begin{array}{cccc}
1 & 0 & 0 & e^{-i\varphi} \\
0 & 1& \pm 1 & 0 \\
0 & \mp 1 & 1 & 0 \\
- e^{i\varphi}& 0 & 0 & 1 \end{array}\right). \en Taking
$\theta=\varphi=0$, the universal quantum gate $\check R$-matrix
(\ref{theorem}) is derived.

\section{The constructions of the Hamiltonian}

 In this section, we present a method of constructing the Hamiltonian
 from the above unitary $\check{R}_\pm(x)$-matrix (\ref{belltype1}) or
 $\check{R}_\pm(\theta)$-matrix (\ref{bellrmatrix}) for the eight-vertex model.

 In terms of the unitary QYBE solution $\check{R}(x)$ (\ref{belltype1}), we construct
 the time-independent Hamiltonian $H_\pm$ having the form \eq H_\pm=i \frac
 {\partial}{\partial x}(\rho_\pm^{-\frac 1 2}\check{R}_\pm)|_{x=1}=- \frac i 2 b^2_\pm =
 \frac i 2 \left(\begin{array}{cccc}
 0 & 0 & 0 & -\,e^{-i\varphi} \\
0 & 0 & \mp 1 & 0 \\
 0 & \pm 1 & 0 & 0 \\
 \,e^{i\varphi}& 0 & 0 & 0 \end{array}\right). \en
Interestingly, we have the ``time-dependent" Hamiltonian
$H_\pm(x)$ by \eq H_\pm(x)=i\,\frac {\partial }{\partial
x}(\rho_\pm^{-\frac 1 2}\check{R}_\pm) \rho_\pm^{-\frac 1
2}\check{R}_\pm^{\dag}(x)=-\frac i {1+x^2}\,\, b^2_\pm \en which
derives the above Hamiltonian $H_\pm$ at $x=1$. When $x$ is real,
the Hamiltonian $H_\pm(x)$ is a Hermition operator.

The wave function $\psi(x)$ is specified by the
$\check{R}(x)$-matrix, with $\psi(x)=\check{R}(x)\psi$, the pure
state $\psi$ independent of the time (or the spectral parameter
$x$). Hence we obtain the Shr{\" o}dinger equation corresponding
to the time evolution of $\psi(x)$ controlled by the
$\check{R}(x)$-matrix, \eq i\,\frac {\partial \psi(x)} {\partial
x}=H(x)\psi(x).\en

For simplicity, we study the unitary $\check{R}(\theta)$-matrix
(\ref{bellrmatrix}) to decide the unitary evolution of quantum
states. After some algebra, the Hamiltonian $H_\pm$ is obtained
 to be  \eq \label{hamiltonian}
 H_\pm=\frac i 2 \frac {\partial  } {\partial \theta}(\rho_\pm^{-\frac 1 2}\check
 R_\pm) \rho_\pm^{-\frac 1 2}
 \check R^\dag_\pm=\frac i 2 \frac {\partial x}{\partial \theta} H_\pm(x)=- \frac i 2 b^2_\pm \en
 which is independent of the time variable $\theta$. So we have
 the expected Schr{\"o}dinger equation \eq i\,\frac {\partial \psi_\pm(\theta)}
{\partial \theta}=H_\pm \psi_\pm(\theta).\en

In terms of the Pauli matrices $\sigma_x$, $\sigma_y$ and $\sigma_z$
and $\sigma_\pm=\frac 1 2(\sigma_x \pm i \sigma_y)$, the Hamiltonian
(\ref{hamiltonian}) has the form \footnote{When the Hamiltonian of
this type was shown to P. Zhang,
 he immediately realized for it another formalism (\ref{hamiltonian2})
 and its relation  (\ref{universal}) to the CNOT gate.}
  \eq \label{hamiltonian1}
H_\pm=\frac i 2\,(-e^{-i\varphi}\sigma_+ \otimes \sigma_+ + e^{i
\varphi} \sigma_-\otimes \sigma_-\mp \sigma_+\otimes \sigma_-\pm
\sigma_-\otimes \sigma_+). \en

Introducing the two-dimensional vector $\vec \sigma$ and two unit
directional vector $\vec n_1$ and $\vec n_2$ in $xy$-plane: \eq
\vec \sigma=(\sigma_x,\sigma_y); \qquad \vec n_1=(\cos\frac
{\pi+\varphi} 2, \sin\frac {\pi+\varphi} 2),\,\, \vec
n_2=(\cos\frac \varphi 2, \sin\frac \varphi 2),  \en the
projections of the vector $\vec \sigma$ into $\vec n_1$ and $\vec
n_2$ are given by \eqa \sigma_{n_1} &=& \vec \sigma \cdot \vec
n_1=\sigma_+ e^{-\frac i 2 (\varphi+\pi)}+\sigma_- e^{\frac i 2
(\varphi+\pi)}, \nonumber\\
\sigma_{n_2} &=& \vec \sigma \cdot \vec n_2=\sigma_+ e^{-\frac i 2
\varphi}+\sigma_- e^{\frac i 2 \varphi}. \ena The Hamiltonian
(\ref{hamiltonian1}) can be recast to \eq \label{hamiltonian2}
H_+=\frac 1 2 \sigma_{n_1}\otimes \sigma_{n_2}, \qquad H_-=\frac 1
2\sigma_{n_2}\otimes \sigma_{n_1}. \en

We consider the unitary time-evolutional operator $U_\pm(\theta)$
determined by the Hamiltonian $H_\pm$, for example, $U_+$ given by
\eq U_+(\theta)=e^{-\frac i 2(\sigma_{n_1}\otimes \sigma_{n_2})
\theta } =\cos\frac \theta 2-\,i\,\sin\frac \theta
2\,\sigma_{n_1}\otimes \sigma_{n_2}. \en

\section{The Bell states and CNOT gate}

 Now we discuss physics related to the time-evolution of the
 universal quantum gate determined by the unitary $\check{R}(\theta)$-matrix
 (\ref{bellrmatrix}). The braid
 group representation $b_\pm(\varphi)$-matrix (\ref{bgrphi}) yields
 the Bell states with the phase factor $e^{-i\varphi}$,
\eq b_\pm(\varphi) \left(\begin{array}{c} |0 0\rangle \\| 0 1 \rangle \\
|1 0\rangle \\ |1 1\rangle
\end{array}\right)=\frac 1 {\sqrt{2}}
 \left(\begin{array}{c} |0 0\rangle-e^{i\varphi}|1 1\rangle \\
 |0 1 \rangle \mp |1 0 \rangle \\ \pm|0 1\rangle + | 1 0 \rangle\\
 e^{-i\varphi} |0 0\rangle+|11\rangle
\end{array}\right)
\en which shows that $\varphi=0$ leads to the Bell states, the
maximum of entangled states, \eq \frac 1 {\sqrt{2}} (|0 0\rangle
\pm |1 1\rangle), \qquad \frac 1 {\sqrt{2}} (| 1 0 \rangle \pm |0
1 \rangle). \en

 In terms of the Hamiltonian $H_\pm$ (\ref{hamiltonian}),
 the $\check R_\pm(\theta)$-matrix (\ref{bellrmatrix}) has the form
 \eq \check{R}_\pm(\theta)=\cos(\frac \pi 4-\theta)+ 2\,i\,
 \sin(\frac \pi 4-\theta)\,H_{\pm}=e^{i (\frac \pi
 2-2\,\theta) H_\pm} \en which can be also used to construct the
 CNOT gate with additional single qubit transformations, examples see
 Ref. \refcite{kauffman8}. The $\check{R}$-matrix (\ref{theorem}) is
 realized as
 \eq
\check{R}=\check R_-(\theta)|_{\theta=\varphi=0}=e^{i\frac \pi 4
(\sigma_x\otimes\sigma_y) }.
 \en

 Applying the unitary $\check{R}(\theta)$-matrix
(\ref{bellrmatrix}) to unentangled states, we have
\eq \check{R}_\pm(\theta) \left(\begin{array}{c} |0 0\rangle \\| 0 1 \rangle \\
|1 0\rangle \\ |1 1\rangle \end{array}\right)=
\left(\begin{array}{l}
 \cos(\frac \pi 4-\theta) |0 0\rangle- e^{i\varphi}\sin(\frac \pi 4-\theta) |1 1\rangle
\\  \cos(\frac \pi 4-\theta) |0 1\rangle \mp \sin(\frac \pi 4-\theta) |1 0\rangle
 \\  \cos(\frac \pi 4-\theta)|1 0\rangle \pm  \sin(\frac \pi 4-\theta) |0 1\rangle)
\\  \cos(\frac \pi 4-\theta) |1 1\rangle + e^{-i\varphi} \sin(\frac \pi 4-\theta)
 |0 0\rangle \end{array}\right). \en Hence with the Bloch vectors on
 the Bloch sphere \cite{nielsen}, the variables $\theta$ and $\varphi$
 realize their geometric meanings so that the construction of the CNOT  gate
 becomes clear.

To obtain other two-qubit quantum gates, for instance the CNOT
gate, we have to apply single qubit unitary transformations $A, B,
C, D$ which can be possibly found in the Bloch sphere
\cite{nielsen} by $SO(3)$ rotations, namely, \eq \label{universal}
 (A\otimes B)U_\pm(\theta)(C\otimes D)=P_\uparrow\otimes
1\!\! 1+ P_\downarrow\otimes\sigma_x=\textrm{CNOT}  \en where the
states $|\uparrow\rangle$ and $|\downarrow\rangle$ are the
eigenvectors of $\sigma_z$,
$\sigma_z|\uparrow\rangle=|\uparrow\rangle,
\sigma_z|\downarrow\rangle=-|\downarrow\rangle$ and the projection
operators $P_\uparrow$ and $P_\downarrow$ have the forms
 \eq
 P_\uparrow=|\uparrow\rangle \langle
  \uparrow|, \qquad P_\downarrow=|\downarrow\rangle\langle
  \downarrow|.
 \en

 Define the $SO(3)$ rotation around the $\vec n$-axis by
 \eq D_{\vec n}(\theta)=e^{-\frac i 2 (\vec \sigma \cdot \vec n)\theta}   \en
 where $\vec \sigma=(\sigma_x, \sigma_y, \sigma_z)$. For examples:
 \eq  D_z(-\frac \varphi 2)=e^{i\frac \varphi 4 \sigma_z}, \,\,
  D_x(\frac \pi 2)=e^{-i\frac \pi 4 \sigma_x}, \,\, D_y(\frac \pi 2)
  =e^{-i\frac \pi 4 \sigma_y} \en
 satisfy
 \eq
 D_x(\frac \pi 2)D_z(-\frac \varphi 2)\sigma_{n_1}D_z(\frac \varphi
 2 ) D_x(-\frac \pi 2)=\sigma_z, \qquad D_z(-\frac \varphi 2)
 \sigma_{n_2}D_z(\frac \varphi 2)=\sigma_x. \en

Consider the evolution operator $U_+(\theta)$. Choosing suitable
single qubit transformations, we obtain \eq ( D_x(\frac
 \pi 2)D_z(-\frac \varphi 2)\otimes D_z(-\frac \varphi 2))U_+
 (\theta)(D_z(\frac \varphi 2)D_x(-\frac \pi 2)\otimes D_z(\frac
 \varphi 2) )= e^{-\frac i 2 (\sigma_z\otimes\sigma_x)\theta}\en
which has another form \eq e^{-\frac i 2
(\sigma_z\otimes\sigma_x)\theta}=P_\uparrow\otimes e^{-\frac i 2
\sigma_x\theta} \, + \, P_\downarrow\otimes e^{\frac i 2
\sigma_x\theta}. \en Set $\theta=\frac \pi 2$. To construct the
CNOT gate, we need additional single qubit transformations
 \eq (\delta\otimes e^{i\frac \pi 4\sigma_x})
 e^{-i\frac \pi 4 (\sigma_z\otimes\sigma_x)}=\textrm{CNOT}
 \en in which the phase gate $\delta$ has the form
 $\delta=P_\uparrow-i\,P_\downarrow$, see (\ref{delta}).

Consider the evolution operator $U_-(\theta=-\frac \pi 2,
\varphi=0)$, namely the $\check{R}$-matrix  (\ref{theorem}) given by
$e^{i\frac \pi 4 (\sigma_x\otimes\sigma_y)}$ which is transformed
into $e^{i\frac \pi 4 (\sigma_z\otimes\sigma_x)}$ by \eq (D_y(-\frac
\pi 2)\otimes D_z(-\frac \pi 2))e^{i\frac \pi 4
(\sigma_x\otimes\sigma_y)}(D_y(\frac \pi 2)\otimes D_z(\frac \pi
2))=e^{i\frac \pi 4 (\sigma_z\otimes\sigma_x)}.\en So we obtain
another proof for {\bf Theorem  1} in the second section or Ref.
\refcite{kauffman8}.

 \section{Concluding remarks}

 Motivated by the observation that there are certain natural similarities
 between quantum entanglements and topological entanglements, we derive
 the unitary solution of the QYBE or the unitary $\check{R}(x)$-matrix
 via Yang--Baxterization and construct the related Hamiltonian for the
 eight-vertex model. With the Brylinksis' Theorem \cite{BB}, the unitary
 $\check{R}(x)$-matrix (\ref{belltype1}) is also a universal
 quantum gate except for $x=1$.

 In the forth-coming article \cite{yong}, we obtain
 more unitary $\check{R}(x)$ solutions determining the evolution of
 universal quantum gates, and we study Yang--Baxterizations of the standard
 or non-standard representations of the six-vertex model and the complete
 solutions of the non-vanishing eight-vertex models.

 \section*{Acknowledgements}

Y. Zhang thanks speakers for helpful lectures in the Fourth
International Wihelm and Else Heraeus Summer School on
``Fundamentals of Quantum Information Processing" in Wittenberg, and
thanks P. Zhang for patiently answering him questions on universal
quantum gates. We thank C.M. Bai and J.M. Hou for helpful
discussions. This work is in part supported by NSFC-10447134.

For L.H. Kauffman, most of this effort was sponsored by the
Defense Advanced Research Projects Agency (DARPA) and Air Force
Research Laboratory, Air Force Materiel Command, USAF, under
agreement F30602-01-2-05022. The U.S. Government is authorized to
reproduce and distribute reprints for Government purposes
notwithstanding any copyright annotations thereon. The views and
conclusions contained herein are those of the authors and should
not be interpreted as necessarily representing the official
policies or endorsements, either expressed or implied, of the
Defense Advanced Research Projects Agency, the Air Force Research
Laboratory, or the U.S. Government. (Copyright 2004.) It gives
L.H. Kauffman great pleasure to acknowledge support from NSF Grant
DMS-0245588.

\end{document}